\documentclass[12pt]{iopart}
\usepackage{graphicx}
\usepackage{dcolumn}
\usepackage{bm}
\usepackage{color}

\newcommand{\CuCu}{{$Cu$+$Cu$\ }}
\newcommand{\XeXe}{{$Xe$+$Xe$\ }}
\newcommand{\PbPb}{{$Pb$+$Pb$\ }}

\newcommand{\hydjet}{{\sc hydjet++ }}

\newcommand{\sqrtsnn}{\sqrt{s_{\mbox{\tiny{\it{NN}}}}}}

\newcommand{\ds}{{\displaystyle}}

\begin{document}

\title{Eccentricities, fluctuations and {\it A}-dependence of elliptic 
and triangular flows in heavy-ion collisions}

\author{G~Kh~Eyyubova$^1$, V~L~Korotkikh$^1$, A~M~Snigirev$^{1,2}$
        and E~E~Zabrodin$^{1,3}$}

\vskip 10mm

\address{$^1$Skobeltsyn Institute of Nuclear Physics, Lomonosov Moscow 
             State University, RU-119991 Moscow, Russia }
\address{$^2$Bogoliubov Laboratory of Theoretical Physics, Joint Institute
             of Nuclear Research, RU-141980 Dubna, Russia }
\address{$^3$Department of Physics, University of Oslo, PB 1048 Blindern,
             N-0316 Oslo, Norway}

\eads{\mailto{evgeny.zabrodin@fys.uio.no}}

\date{Received: date / Accepted: date}
\begin{abstract}
A simple geometrical model with event-by-event fluctuations is suggested
to study elliptical and triangular eccentricities in the initial state
of relativistic heavy-ion collisions. This model describes rather well
the ALICE and ATLAS data for Pb+Pb collisions at center-of-mass energy
$\sqrtsnn = 5.02$~TeV per nucleon pair, assuming that the second, $v_2$, 
and third, $v_3$, harmonics of the anisotropic flow are simply linearly 
proportional to the eccentricities $\varepsilon_2$ and $\varepsilon_3$, 
respectively. We show that the eccentricity $\varepsilon_3$ has a pure 
fluctuation origin and is substantially dependent on the size of the 
overlap area only, while the eccentricity $\varepsilon_2$ is mainly 
related to the average collision geometry.
Elliptic flow, therefore, is weakly dependent on the event-by-event
fluctuations everywhere except of the very central collisions 0--2\%,
whereas triangular flow is mostly determined by the fluctuations. The
scaling dependence of the magnitude of the flow harmonics on atomic
number, $v_n \propto A^{-1/3}$, is predicted for this centrality
interval.
\end{abstract}

\pacs{25.75.-q, 24.10.Nz, 24.10.Pa} 


\maketitle


\section{Introduction}
\label{intro}

Study of properties of new state of matter, the quark-gluon plasma (QGP),
is one of the main goals of experiments on relativistic heavy-ion 
collisions carried out at Relativistic Heavy Ion Collider (RHIC) and
Large Hadron Collider (LHC), and planned ones at coming in the nearest
future Nuclotron-based Ion Collider fAcility (NICA) and Facility for
Antiproton and Ion Research (FAIR); see 
\cite{qm_19,qm_18} and references therein. 
Among the signals most sensitive to the formation of QGP and its
subsequent hadronization is the phenomenon known as collective flow;
for reviews see, e.g., \cite{ReRi_97,HWW_99,VPS_10,HeSn_13}.
It employs a Fourier series expansion of hadron distribution in the
azimuthal plane \cite{VoZh_96,PoVo_98}
\begin{equation}
\ds
\label{eq:1}
\frac{dN}{d\varphi} \propto 1 + 2\sum\limits_{n = 1}^\infty v_n
\cos{ \left[ n(\varphi -\Psi_n) \right] } \ .
\end{equation}
The infinite series in the r.h.s. of Eq.(\ref{eq:1}) represent the 
anisotropic flow. The latter contains the flow harmonics $v_n$. 
$\varphi$ is the azimuthal angle of transverse momentum of particles 
in the laboratory frame and $\Psi_{n}$ is the symmetry plane of the 
$n$th harmonic also in the laboratory frame. 
The flow coefficients are calculated as
\begin{equation}
\label{eq:2}
v_n =  \langle\langle \cos{ \left[ n(\varphi -\Psi_n) \right]}
\rangle\rangle \ ,
\end{equation}
where the averaging is taken over all hadrons in a single event and over 
all measured events. The first harmonics are dubbed directed, $v_1$,
elliptic, $v_2$, triangular, $v_3$, quadrangular, $v_4$, flows, and so 
forth. Present paper deals with the study of centrality dependencies of 
$v_2$ and $v_3$.

Elliptic flow has been intensively studied both theoretically
\cite{Oll_92,Oll_93,Sor_99,plb_01,Tea_03,MoVo_03,SJG_11,Rug_13,Fev_18}
and experimentally
\cite{ST_01,PH_03,AL_10,AT_12,CMS_12,ST_13}
during the last three decades. It comes from the transformation of the
initial anisotropy of the overlap region in the coordinate space into 
the momentum anisotropy of expanding fireball. In contrast to $v_2$, 
triangular flow in heavy-ion collisions is drawing attention in the
last decade only \cite{AlRo_10,ACLO_10,AL_11,SJG_12}. 
For a quite long period it was believed that the odd
harmonics of anisotropic flow, such as $v_3$, $v_5$, and so on, vanishes
in collisions of similar nuclei because of the symmetry considerations.
However, as was first shown in \cite{AlRo_10}, fluctuations in initial
state 
(mainly, the nucleon positions) cause the triangular anisotropy 
$\varepsilon_3$ of the
collision region. Triangular flow, therefore, possesses its own 
symmetry plane $\Psi_3$, which is randomly oriented w.r.t. the
position of the symmetry plane $\Psi_2$ of elliptic flow. 
Among the
interesting features of both $v_2$ and $v_3$ is their approximately
linear dependence on the corresponding eccentricity, $\varepsilon_2$
and $\varepsilon_3$ \cite{Vel_11,Jia_11,Sn_11,Pet_10},  
and their significant contribution to higher flow harmonics
\cite{GGLO_12,TY_12,prc_14}. 

Having the LHC put into operation, one got access to a number of 
experimental intriguing and exquisite phenomena which would have never 
been systematically studied at the accelerators of previous generations. 
In this paper we explore and draw attention to the centrality 
dependence of elliptic and triangular flows. Such dependence has been 
thoroughly measured by the 
ALICE~\cite{ALICE_2018}, 
ATLAS~\cite{ATLAS_2012,ATLAS_2019} and CMS \cite{CMS_2014} 
Collaborations in \PbPb collisions at center-of-mass energy 2.76~TeV 
and 5.02~TeV per nucleon pair. 
In particular, these measurements 
demonstrate the nontrivial centrality dependence of the ratio of the 
second flow harmonic to the third one, which is typically missed in 
calculations~\cite{Alba:2017hhe} based on hydrodynamic scenario of 
the fireball evolution and in many phenomenological 
approaches; for instance, in the popular \hydjet model 
\cite{Lokhtin:2008xi,Bravina:2013xla,prc_09,prc_21}. This 
observation has attracted a lot of attention now; see, e.g., 
\cite{Gelis,Floerchinger:2020tjp,Zakharov} and references therein. 
To investigate the centrality dependence of elliptic and triangular 
flows a simple geometrical model with event-by-event fluctuations is 
considered here. 

The paper is organized as follows. The basic principles of the 
approach are described in Sec.~\ref{sec_2}. Section~\ref{sec_3} presents
the numerically calculated centrality dependencies of $v_2$ and $v_3$ in
lead-lead collisions at $\sqrtsnn = 5.02$~TeV. A fair agreement with the
experimental data is obtained. In particular, we show that the profile
of the overlap area, circular or elliptical, in the transverse plane is 
almost unimportant for the formation of triangular flow compared to the 
initial-state fluctuations. Dependence of both, $v_2$ and $v_3$, in very 
central collisions on atomic number of colliding nuclei is discussed. 
Finally, conclusions are drawn in Sec.~\ref{concl}.


\section{Formalism and modeling}
\label{sec_2}

The standard procedure to determine the eccentricities and the 
effective overlap area of two nuclei colliding with impact parameter 
{\bf b} employs Glauber eikonal model \cite{Gl_59,MRSS_07}. 
For recent review describing the basic formalism see, e.g., 
\cite{dEnterria:2020dwq}. To define the ``reaction centrality'' $C$,
the cross section of particle production is subdivided into centrality 
bins $C_k=C_1, C_2, \ldots$. The width of each bin $\Delta C$ 
corresponds to some fraction part of the total cross section. For 
instance, typical choice for most central collisions is 5\% meaning 
that $\Delta C = 0.0 - 0.05$. Collision of two similar nuclei with 
radii $R$ in this model corresponds to collision of two black disks 
of the same radii with centrality $C = b^2/(4R^2)$, providing us
$C = 100\%$ for $b = 2R$.

Then, the thickness function $T_A(x,y)$ is the main quantity of 
interest in Glauber approach
\begin{equation}
T_A(x,y) = \int dz  \; \rho_A(x,y,z)\ ,
\label{eq:Tp2}
\end{equation}
where the three-dimensional nuclear density $\rho_A(x,y,z)$ is 
determined via the standard Fermi-Dirac, or rather Woods-Saxon, 
distribution 
\begin{equation}
\ds
\rho_A(x,y,z)=\rho_0 \; \frac{1}{ e^{(r-R)/ d} +1}\ .
\label{eq:Tp3}
\end{equation}
Here $R$ is the nuclear radius, $A$ is its atomic number, $d$ is the 
diffuseness edge parameter and $\rho_0$ is a normalization constant, 
so that $\int \rho_A(r) \; d^3r = A$. 

The needed eccentricities are calculated by the standard formulas
\begin{eqnarray}
\ds
\varepsilon_n &=& \varepsilon_{n,x} + i\varepsilon_{n,y} = 
\frac{\int s \; ds \; d\phi \; e^{i n\phi} s^n w ({\bf s, b}) }
{\int s \; ds \; d\phi \; s^n w({\bf s, b})}\ ,\nonumber\\
|\varepsilon_n|^2 &=&  \varepsilon_{n,x}^2 + \varepsilon_{n,y}^2\ ,
\label{eq:Tp9}
\end{eqnarray}
where $w({\bf s, b})$ are some weights, $s^2=x^2+y^2$, and
$\tan\phi =y/x$.
Under the assumption of some ``macroscopic'' overlap between the 
colliding nuclei one can define the transverse overlap area as 
\cite{dd2010}
\begin{equation}
S(b) = 4\pi \sqrt{\langle x^2 \rangle \, \langle y^2 \rangle}\ ,
\label{eq:Tp10}
\end{equation}
where the weighted averages are the same as in Eq.(\ref{eq:Tp9}).
Recall that one lacks unique definition of the absolute normalization 
of the overlap area. In our definition the overlap area has maximum 
magnitude $4\pi$, which is four times larger than, e.g., that defined
in Ref.~\cite{VoZh_96}. On the other hand, it almost coincides with 
the geometrical overlap area between two disks possessing uniform 
two-dimensional density distribution.    

The next step is a choice of weights, which is ambiguous and is 
determined in the specific models. Often the number of binary 
nucleon-nucleon collisions in an $A+B$ collision at given impact 
parameter ${\bf b}$ is used as weights. In this case up to the 
inessential normalization factor they are equal to
\begin{equation}
w({\bf s, b}) = T_A(x+b/2,y) T_B(x-b/2,y)
\label{weights}
\end{equation}
with the impact parameter ${\bf b}$ being directed along the $x$-axis. 
Other weights are possible too, for instance, the number of 
participating nucleons. This problem was partly discussed in 
\cite{dd2010} with calculations of second eccentricity $\varepsilon_2$ 
for various nucleon densities as a function of $b$. In particular, at a 
small value of diffuseness edge parameter $d$ the eccentricity 
$\varepsilon_2$ is close to the pure geometrical one
\begin{equation}
\varepsilon_{2,\rm geom}=b/(2R) =\sqrt{C} \ . 
\label{e2-g}
\end{equation}
In this approach the third eccentricity $\varepsilon_3$, as well as the
other odd ones, is identically equal to zero if the three-dimension 
nuclear density $\rho_A(x,y,z)$ depends on the radial distance $r$ 
only (a spherically symmetric distribution of matter in the Breit 
system).

Here one should note that nuclei consist of a finite number of 
nucleons which have the finite sizes and their positions are 
distributed in accordance with Eq.(\ref{eq:Tp3}). 
The nucleon positions can fluctuate in event-by-event. These 
fluctuations were taken into account in Monte Carlo Glauber 
calculations~\cite{MRSS_07} and subsequently with the realization that 
odd flow coefficients would be non-zero~\cite{AlRo_10}. The energy 
deposition in the overlap region of two nuclei can also be used as 
weights taking into account its fluctuations 
\cite{Gelis,Floerchinger:2020tjp}. Both approaches are discussed and 
reviewed in the subsections below.


\subsection{Geometrical model with fluctuations}
\label{subsec_2a}

In the simplest model of hard spheres the nuclei have a spherical shape 
of radius $R$ with uniformly distributed nucleon density
\begin{equation}
\rho_A(x,y,z)= \frac{3}{4\pi R^3} \; \Theta (r-R).
\label{hard-sphere}
\end{equation}
In this case the eccentricities and the effective overlap area in the 
non-central collision of two nuclei with the impact parameter {\bf b} 
are calculated in an analytic form that are often used as a first good 
approximation. 

The model can be improved by taking into account the fluctuations of 
nucleon positions. Note that the integration in Eq.(\ref{eq:Tp9}) can 
be performed by the Monte Carlo method. The accuracy of estimation is 
limited by the number of points in which weights are calculated. 
Therefore, we opted to calculate the integrals in Eq.(\ref{eq:Tp9}) 
by the Monte Carlo method with the {\it finite fixed number} of points 
$M$ over each of coordinate, $x$ and $y$. In this approach a single
event represents the location of $M^2$ points in the $x$-$y$ plane. 
At this simple modeling the number $M^2$ is a some analogue of the 
number of the participant nucleons or colour charges used in other 
models.
Then, one should average over ensemble of such events with the fixed 
finite number of points. As a result of averaging procedure
$\langle\varepsilon_{3,x}\rangle = \langle\varepsilon_{3,y}\rangle = 0$, 
but
\begin{equation}
\varepsilon_3\{2\} = \sqrt{\langle |\varepsilon_3|^2 \rangle}
\end{equation}
will be non-zero as well as 
\begin{equation}
\varepsilon_2\{2\} = \sqrt{\langle |\varepsilon_2|^2 \rangle}
\end{equation}
at $C = 0$ in contrast to ``continuous'' ($M \rightarrow \infty$) 
calculations (\ref{e2-g}). The number of points, in which weights are 
calculated at $C = 0$, is a single parameter which should be fitted to 
reproduce data.


\subsection{Magma model}
\label{subsec_2b}

The resulting profile of energy density in an ultrarelativistic 
nucleus-nucleus collision is simulated as the sum of 
contributions of elementary collisions between a localized color charge 
and a dense nucleus. Each elementary collision yields a source of energy 
density which is independent of rapidity and decreases with distance 
from the center of the source. Thus, the energy density 
$\rho({\bf r,b})$ as a function of the transverse distance ${\bf r}$ 
and the impact parameter ${\bf b}$ is determined by the product of the 
saturation momentum squared $Q^2$ of one nucleus 
(where $1/Q^2$ is 
related to an effective ``area" of each colour charge in the transverse 
region)
 and the random source depositions of other nucleus \cite{Gelis}
\footnote{We are aware that the Magma authors pointed out a potential 
problem with the CGC correlator employed in the model. This correlator 
is not directly used in our study.} 
\begin{eqnarray}
\rho({\bf r,b}) &=& \sum_{j\subset A} Q^2_B({\bf s}_{A,j},{\bf b})
       \Delta_A ({\bf r}- {\bf s}_{A,j},{\bf b}) \nonumber\\
&+& \sum_{j\subset B} Q^2_A({\bf s}_{B,j},{\bf b})
       \Delta_B ({\bf r}- {\bf s}_{B,j},{\bf b})\ .
\label{eq:Tp7}
\end{eqnarray}
Here $Q_A$ and $Q_B$ are the saturation momenta of the colliding nuclei 
$A$ and $B$ to be specified below. The positions ${\bf s}_{A,j}$ and 
${\bf s}_{B,j}$ are assumed to be independent random variables. 

The profile $\Delta$ of energy source in nucleus ($A/B$) is selected 
in the form which satisfies the short distance correlations in the 
Color Glass Condensate (CGC) approach \cite{cgc1,cgc2,cgc3}
\begin{equation} 
\ds
\Delta ({\bf r}- {\bf s}_{A,j},{\bf b})
 =  \left\{ \begin{array}{ll}
\frac{8}{g^2 N_c}\frac{1}{|{\bf r} - {\bf s}_{A,j}|^2 + 
   Q^{-2}_A( {\bf r},{\bf b}) }, & |{\bf r}- {\bf s}_{A,j}|<1/m\\
0, & |{\bf r}- {\bf s}_{A,j}|>1/m
\end{array} \right\}\ ,
\label{eq:Tp5}
\end{equation}
where $g$ is the dimensionless coupling constant of QCD, $N_c$ is the 
number of colors ($N_c=3$ for QCD), and $m$ is the infrared cutoff 
parameter of the order of pion mass. At large distance  
$\Delta{\bf (r)}$ decreases like $1/r^2$ as for a Coulomb field in 
two dimensions. However, $\Delta{\bf (r)}$ goes to a finite value for 
$r \rightarrow 0$, while it would diverge for a pointlike charge. The 
physical interpretation is that the charge is spread over a distance 
$\sim 1/Q_A$. The number of elementary charges contained in an area of 
this size is of the order $1/g^2$, which explains the normalization 
factor arising in Eq.(\ref{eq:Tp5}).

If a source is located in the region of the nuclear size, then the 
distribution (\ref{eq:Tp5}) is concentrated inside an area with radius 
$|{\bf r}| < 1/m = 1.4$ fm. It is considerably smaller than the 
transverse area of heavy nuclei (for instance, for lead nucleus with 
radius $ R=6.62$ fm). At $ {\bf s}_{A,j} = {\bf r} $ the energy 
intensity reaches maximum in the nuclear center and is proportional to
$Q^{2}_A( {\bf r,b})$. The integral intensity of one source, to the 
leading logarithm accuracy, reads
\begin{equation}
\ds
I_A({\bf r,b}) = \int d^2s \Delta ({\bf r}- {\bf s},{\bf b})
\simeq \frac{8 \pi}{g^2N_c} 
\ln{ \left[ 1+\frac{Q^{2}_A({\bf r,b})}{m^2} \right]}\ .
\label{eq:Tp6}
\end{equation}

In the Magma model $Q^2_A$ is assumed to be proportional 
to the integral of the nuclear  density over the longitudinal 
coordinate $z$, i.e., to the thickness function
\begin{equation}
Q^2_A(x,y) = Q^2_{s0}T_A(x,y)/T_A(0,0)\ .
\label{eq:Tp1}
\end{equation}
The value of the saturation momentum $Q_{s0}$ at the nucleus center 
is a free parameter in this approach and the energy density 
$\rho({\bf r,b})$ is applied as weights in calculations of 
eccentricities (\ref{eq:Tp9}).

The number of sources per unit area (density) is eventually given by
\begin{equation}
\ds
n_A  = \frac{N_c^2}{32\pi}\frac{Q^2_A({\bf r,b})}
{\ln{ \left[ 1+\frac{Q^{2}_A( {\bf r,b})}{m^2 } \right]}}\ .
\label{eq:Tp8}
\end{equation}
Thus, the maximum number of sources in the nuclear transverse area 
with radius $R=6.62$~fm at $Q_{s0} = 1.24$~GeV is approximately 
calculated as $N_ {\rm Pb} \simeq 100$. Note, that the parameter 
$1/\sqrt{N_{\rm Pb}}$ can characterize the fluctuation scale. 


\subsection{(Improved) Monte Carlo Glauber model}
\label{subsec_2c}

The Monte Carlo (MC) Glauber model \cite{Gl_59,MRSS_07,dEnterria:2020dwq}
is widely used in analysis of experiments with relativistic heavy-ion
collisions. It relates the initial collision geometry to the measured
observables. Its basic principles have been already discussed in the
beginning of this Section. Recall, that the MC Glauber model
treats nucleus-nucleus collisions as a superposition of independent
nucleon-nucleon collisions. The positions of nucleons relative to the
geometrical centers of colliding nuclei are assumed to be randomly 
distributed in space according to Woods-Saxon spherical density 
$\rho(\vec{r})$, see Eq.(\ref{eq:Tp3}). Quite often, this distribution 
is taken from the available experimental data. Nucleons interact if the 
distance between their centers is less than $\sqrt{\sigma_{inel}^{NN} /
\pi}$, where $\sigma_{inel}^{NN}$ is the inelastic cross section of
$NN$ interaction. 
In eikonal approximation nucleons move along straight-line trajectories. 
To improve the modeling, it was also
suggested to employ $NN$ correlations \cite{ADS_09}.

In the present paper we use the improved MC Glauber model described in
\cite{LKE_18}. In contrast to standard MC Glauber model, the improved
model implements separated transverse profiles for neutrons and protons
in heavy nuclei, such as gold or lead ones. The positions of nucleons 
inside a nucleus are modeled to provide a minimum separation in order
to emulate hard-core repulsion between nucleons. The latter should not, 
however, distort the nuclear density.

It is worth mentioning that in MC Glauber model the number of 
participants serves as weights for eccentricity calculations. Therefore,
all three models at our disposal rely on different weighting schemes:
MC hard sphere model uses the number of binary collisions, Magma model
employs the energy density, and MC Glauber model applies the number of 
participants. In the next Section we will see how these differences 
affect the estimations of elliptic and triangular eccentricities and 
related to it elliptic and triangular flows in heavy-ion collisions.


\section{Results}
\label{sec_3}

\begin{figure}[htpb]
\begin{center}
\resizebox{0.8\textwidth}{!}{%
\includegraphics{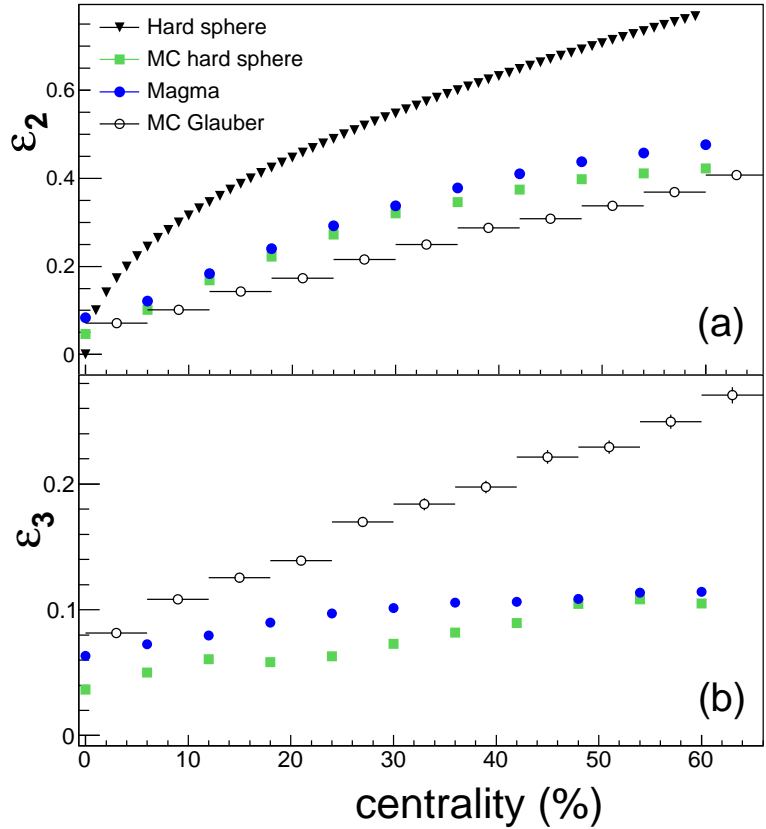}}
\end{center}
\caption{(Color online)
The eccentricities (a) $\varepsilon_2\{2\}$ and (b) $\varepsilon_3\{2\}$ 
as functions of centrality $C$ for hard sphere
model (triangles), Monte Carlo hard sphere model (squares), Magma
model (full circles), and Monte Carlo Glauber model (open circles). }
\label{fig1}
\end{figure}

The formalism briefly reviewed in Sec.~\ref{sec_2} is applied to 
calculate the initial eccentricities $\varepsilon_2$ and $\varepsilon_3$ 
as functions of the geometrical centrality $C$ in relativistic heavy-ion 
collisions. 
Values accessible experimentally are moments or 
cumulants of the distribution of the flow harmonic coefficients $v_n$. 
The lowest order cumulants are defined as~\cite{borghini}
\begin{eqnarray}
v_n\{2\} &=& \sqrt{\langle |v_n|^2 \rangle}\ .
\label{eq:vn}
\end{eqnarray}
In the linear response approximation they are simply proportional to 
the corresponding cumulants of the initial eccentricities
\cite{GGLO_12}
\begin{eqnarray}
v_2\{2\} &=& k_2 \varepsilon_2\{2\}\ , \nonumber \\
v_3\{2\} &=& k_3 \varepsilon_3\{2\}\ .
\label{eq:vn-e2}
\end{eqnarray}
Typically, the cumulants are measured over the two-particle 
correlations defined as
\begin{equation}
\langle\langle 2\rangle\rangle =  
      \langle\langle e^{in(\varphi_1-\varphi_2)}\rangle\rangle =
      (v_n\{2\})^2 \ .
\end{equation}
Here the double averaging  $\langle\langle \rangle\rangle$ is performed 
over all particle combinations and over all events; see, e.g., 
\cite{prc_21} for the definition of the differential and other 
cumulants over particle correlations.
For the collisions with centrality from 0\% to 30\% the Magma model 
describes successfully~\cite{Gelis} the experimental data on $v_2$ and 
$v_3$ as functions of centrality percentile, measured by the ATLAS 
Collaboration~\cite{ATLAS_2019} in \PbPb collisions at $\sqrtsnn = 
5.02$~TeV. The proportionality coefficients $k_2=0.321$ and 
$k_3 =0.314$ together with the saturation momentum $Q_{s0}=1.24$~GeV 
were adjusted to data. 
This set of parameters was employed to calculate the
centrality dependence of the eccentricities $\varepsilon_2$ and
$\varepsilon_3$. Obtained results are compared with predictions of
other models, such as Monte Carlo hard sphere, hard sphere, and Monte 
Carlo Glauber model, in Fig.~\ref{fig1}. Recall that for 
calculations we used the improved Monte Carlo Glauber model, described 
in \cite{LKE_18}. Cross section $\sigma_{NN} = 67.6$~mb was picked up 
from Table~III of this paper. 

In Fig.~\ref{fig1}(a) one can see that the 
elliptic eccentricity $\varepsilon_2$ calculated in the Magma model is
very close to that obtained in the Monte Carlo hard sphere model within 
the centrality interval 2--30\%. Note, that for the very central 
collisions with $\sigma/\sigma_{geo} = 0-2\%$ both Magma and MC hard 
sphere eccentricities go to a finite value, whereas the geometrical 
$\varepsilon_2 = b/(2R)$ goes to zero at $C \rightarrow 0$.
Calculations of MC Glauber model are qualitatively similar to those of
two other MC models (MC Hard sphere, Magma) but lie about $20-30\%$ 
below. For the triangular eccentricity, shown in Fig.~\ref{fig1}(b), 
the Magma model overpredicts the MC hard sphere results by 25--40\% 
for centralities $C \leq 40\%$. For more peripheral collisions 
predictions of both models quickly converge. 
The MC Glauber model significantly overpredicts the calculations of 
both, MC hard sphere and Magma, models. Recall, that the odd-order 
eccentricities are zero in the hard sphere (HS) model, therefore, 
$\varepsilon_3^{HS}$ is absent in this figure.

\begin{figure}[htpb]
\begin{center}
\includegraphics[width=7.00cm]{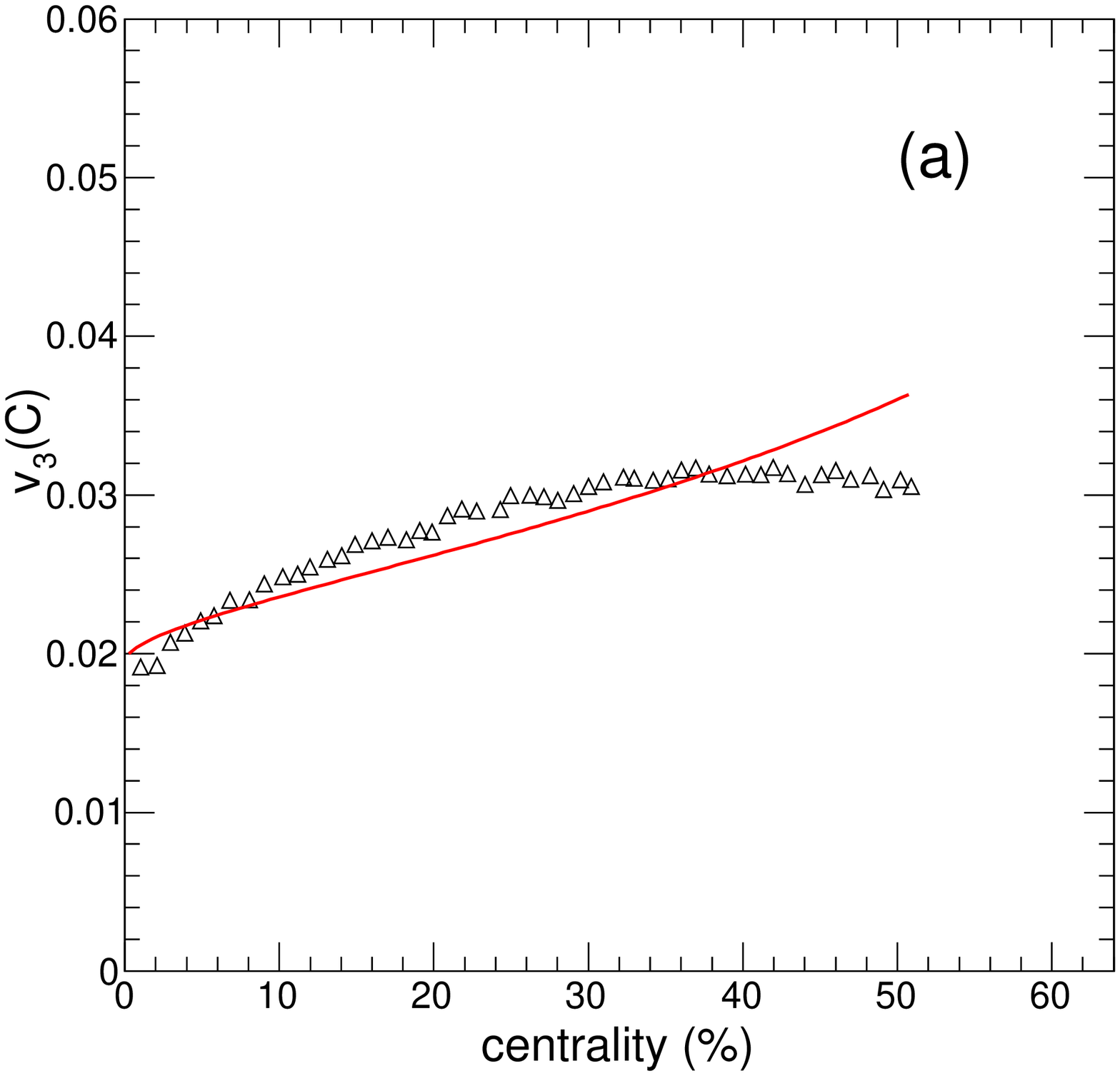}
\includegraphics[width=7.00cm]{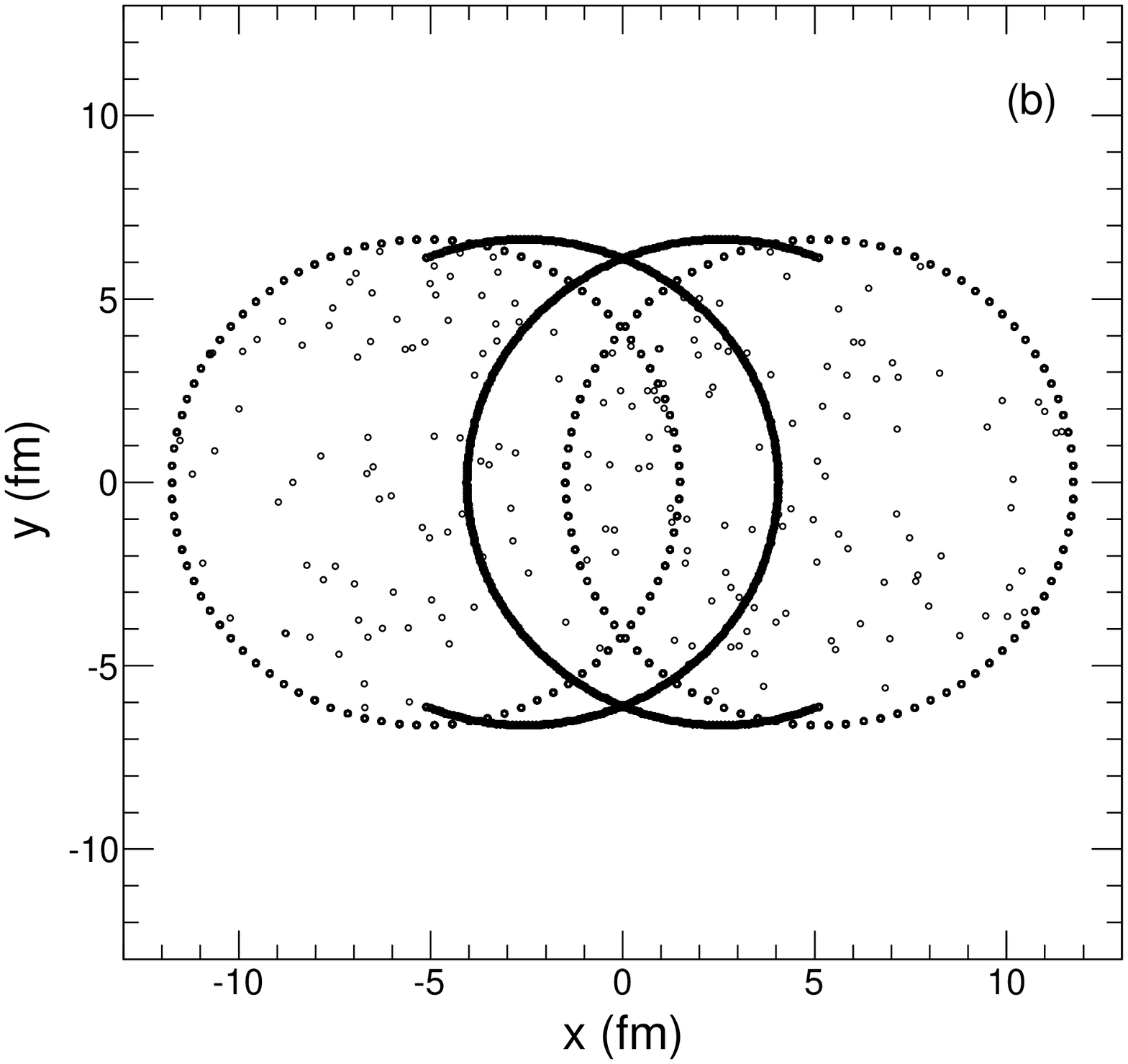} \\
\includegraphics[width=7.00cm]{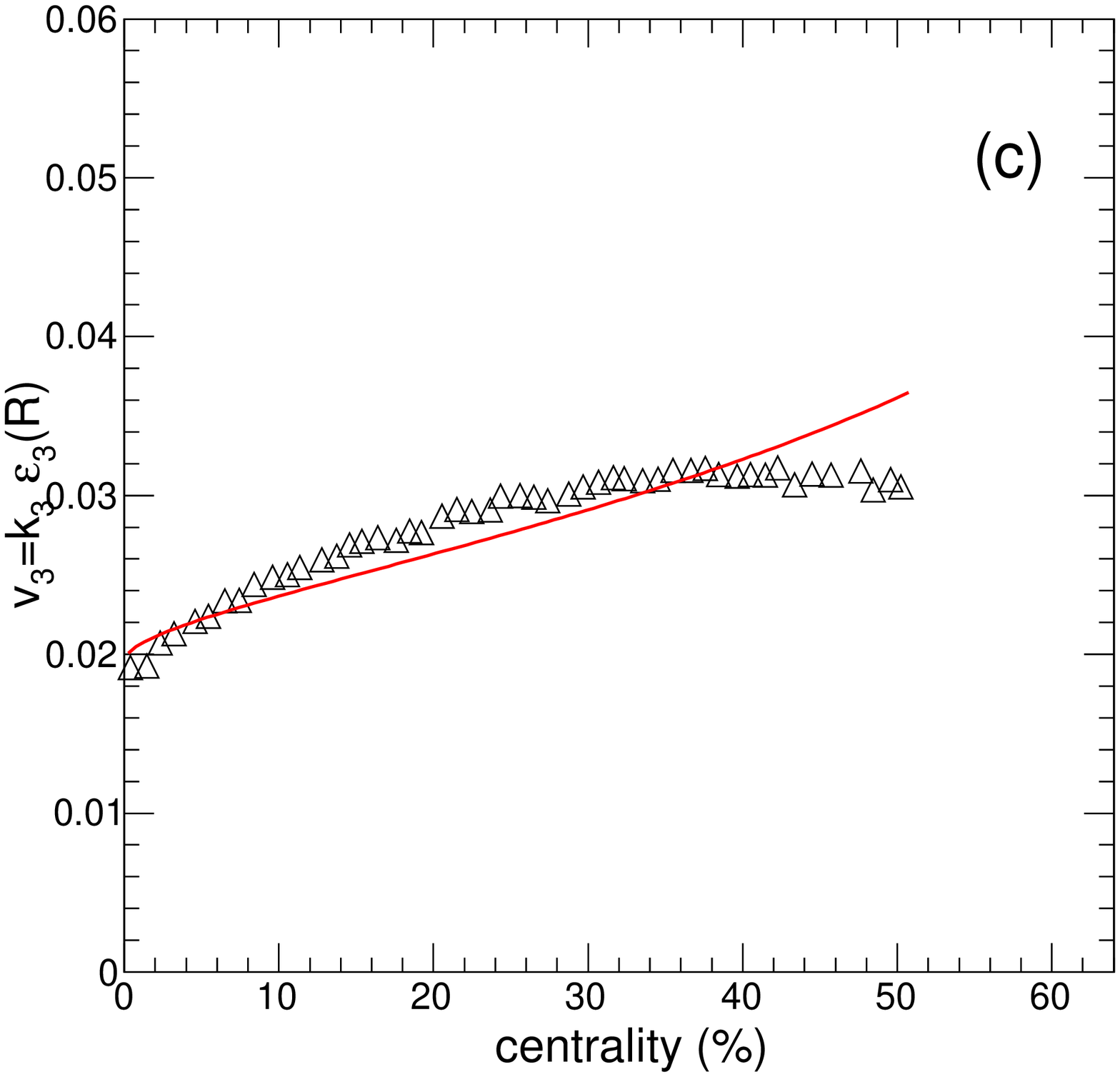}
\includegraphics[width=7.00cm]{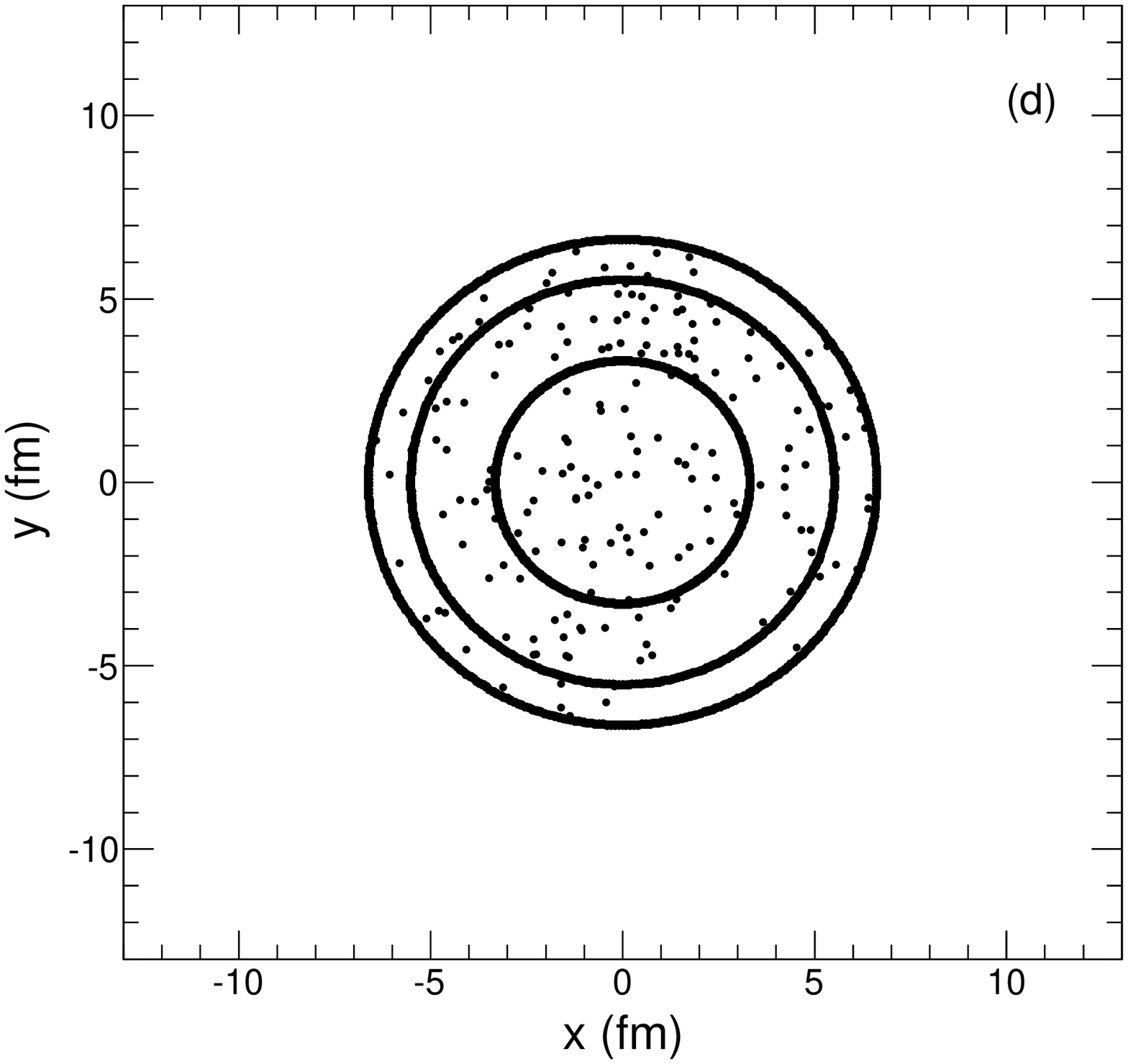}
\end{center}
\caption{(Color online)
(a) Centrality dependence of triangular flow in \PbPb collisions at
$\sqrtsnn = 5.02$~GeV. Solid line shows the calculations in accordance 
with Eq.~(\ref{eq:Tp11}), whereas open triangles denote the ALICE data
from \cite{ALICE_2018}
in $Pb+Pb$ collisions at center-of-mass
energy 5.02 TeV per nucleon pair within the transverse momentum
interval $ 0.2 < p_{T} < 50$ GeV/$c$ and within the pseudorapidity
interval $|\eta|< 0.8$. A separation in pseudorapidity between the
correlated particles $|\Delta\eta|>1$ is applied. All observables are
calculated in small centrality bins (1\%).
(b) Changing of the overlap region in transverse plane with variation 
of the collision centrality from $C = 12\%$ (solid curves) to 60\% 
(dashed curves). Dots show the positions of the sources. See text for 
details.
(c) Triangular flow calculated in both models (solid curve) at $b=0$ 
with the radius variation in accordance with Eq.~(\ref{eq:Tp12}) and at 
$k_3 = 0.314$ in Eq.~(\ref{eq:vn-e2}). The radius of the circle is 
varying to provide the same area as that of the overlap region of 
\PbPb collisions at certain centrality.
Open triangles denote the ALICE data from \cite{ALICE_2018}. 
(d) Changing of the overlap area with the radius variation from
$R = 6.62$~fm (outer circle) to 5.3~fm (inner circle) and 3.1~fm (most
inner circle). Dots show the positions of the sources. See text for 
details.}

\label{fig2}
\end{figure}

In contrast to the second eccentricity coefficient determined by the 
shape of the overlap region and, therefore, mainly by the collision 
geometry, the third eccentricity coefficient is of a pure fluctuation 
origin. 
Its magnitude is practically  determined by the overlap region area, 
but not its shape. The indirect evidence of this affirmation follows 
from the fact that the ratio $\sqrt{\langle v^2_3 \rangle - \langle
v_3 \rangle^2}/ \langle v_3 \rangle $ is nearly constant at all 
centralities and is slightly dependent on the transverse momentum.
The sensitivity to the overlap shape can be tested by the variation
of the diffuseness edge parameter $d$. We found that, unlike the 
second eccentricity $\varepsilon_2$ which is noticeably dependent of 
$d$, the third eccentricity is practically insensitive to the
diffuseness variation, i.e., to the shape of overlap region.
Figure \ref{fig2} illustrates these statements. 
Indeed, in the simple 
model of hard sphere one can easily calculate the area of the overlap 
region $S(C)\simeq \pi R^2 (1-\sqrt{C})$ and, therefore, the number of 
``source-points'' $N_{\rm points} = {\rm density} \times S(C)$. If the 
third harmonic has a pure fluctuation origin then its magnitude reads
\begin{equation}
\ds
v_3(C) \sim \frac{1}{\sqrt{N_{\rm points}}} \simeq \frac{K_3}
{\sqrt{ 1 - C^{1/2}}  }\ .
\label{eq:Tp11}
\end{equation}
This pure fluctuation centrality dependence~(\ref{eq:Tp11}) is shown 
in Fig.~\ref{fig2}(a),(b) in comparison with experimental data. With 
$K_3=0.0183$ we obtain a good description of the data in the 
centrality region below 50\%.

To demonstrate that the area of the overlap region has the significance 
only, while the shape is unimportant, we display in 
Fig.~\ref{fig2}(c),(d) 
the triangular flow $v_3$ calculated for perfectly central collisions
with $b=0$, but with the radius changing as
\begin{equation}
\ds
R (C) = R \sqrt{1 - C^{1/2}} \ .
\label{eq:Tp12}
\end{equation}
In this case the overlap region has circular shape, see 
Fig.~\ref{fig2}(d). However, the area of the circle is the same as 
the overlap area in the hard sphere model at given centrality $C$. 
As one can see in Fig.~\ref{fig2}(c), a good agreement with 
the experimental data is obtained despite of the fact that no new
parameters were either introduced or tuned in the model.

\begin{figure}[htpb]
\begin{center}
\resizebox{0.8\textwidth}{!}{%
\includegraphics{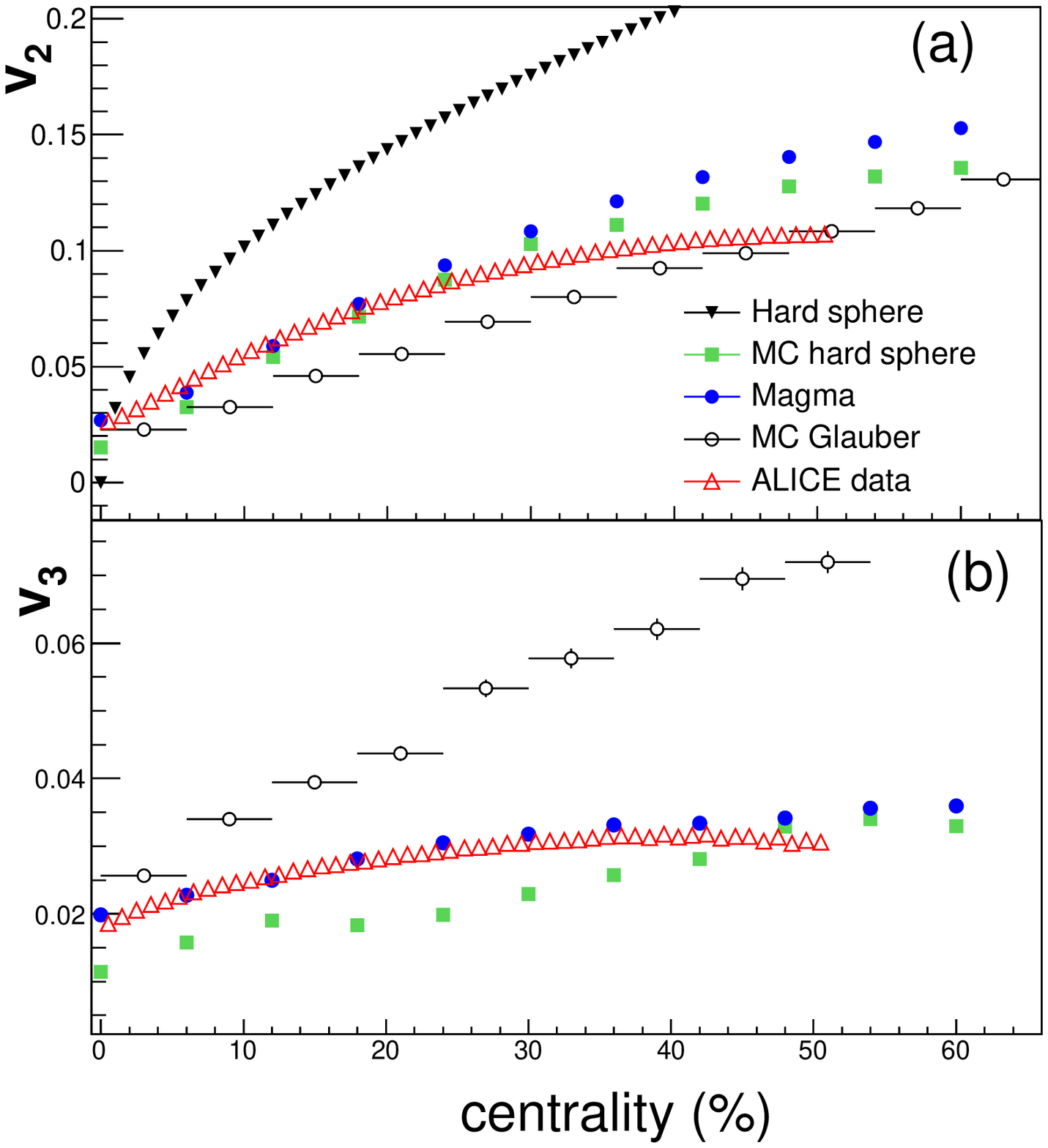}}
\end{center}
\caption{(Color online) 
The elliptic $v_2\{2\}$ and triangular $v_3\{2\}$ flows as functions of 
centrality $C$ for hard sphere model (full triangles),
Monte Carlo hard sphere models (full squares),
Magma model (full circles),
and MC Glauber model (open circles). 
Open triangles denote the ALICE data from \cite{ALICE_2018} 
measured in $Pb+Pb$ collisions at center-of-mass
energy 5.02 TeV per nucleon pair within the transverse momentum
interval $ 0.2 < p_{T} < 50$ GeV/$c$ and within the pseudorapidity
interval $|\eta|< 0.8$. A separation in pseudorapidity between the
correlated particles $|\Delta\eta|>1$ is applied. All observables are
calculated in small centrality bins (1\%).}
\label{fig3}
\end{figure}

Figure~\ref{fig3} shows the elliptic $v_2\{2\}$ and triangular 
$v_3\{2\}$ flows, calculated by two-cumulant method \cite{borghini} 
as functions of centrality $C$ in the linear response approximation,
given by Eq.(\ref{eq:vn-e2}), with the {\it constant} coefficients 
adjusted to data. We see that for more peripheral collisions with 
centralities larger than 30\% a linear dependence $ v_2(C)=k_2 
\varepsilon_2(C)$ with constant coefficients $k_2$ is not realized, 
since the second eccentricity $\varepsilon_2\{2\}$ becomes too large.
The $k_2$ ``hydro'' conversion coefficient is expected to change with 
centrality, smaller in peripheral collisions,
and nonlinear effects come into play.

The results of MC Glauber model for $v_2(C)$, presented in 
Fig.~\ref{fig3}(a), demonstrate weaker elliptic flow compared to that of
two other MC models in the centrality range $5\% \leq C \leq 50\%$. The 
flow measured by the ALICE Collaboration lies between the three model
calculations. It implies that elliptic flow is mainly determined by the
geometry of the overlap region, whereas fluctuations play a minor role
for all but very central collisions. For the triangular flow, displayed
in Fig.~\ref{fig3}(b), the fluctuations induced by the MC Glauber model 
are too strong compared to data, leaving room for further tuning of 
model parameters. 

The ``pure fluctuation'' contribution to the second harmonic reveals 
itself in the most central collisions with $\sigma/\sigma_{geo} = 0-2\%$
only, where the ``pure geometrical'' contribution goes to zero. This 
explains also the interesting observation that the ratio $v_2/v_3 = 1$ 
at $C=0$. One should note that the absolute magnitude of both harmonics 
within this centrality interval is merely determined by the number of
sources, i.e., by the area of overlap region. It means that MC models 
predict the larger values of elliptic and triangular flows in very 
central collisions for the lighter nuclei, because $v_2$ and $v_3$ 
turn out to scale with the atomic number to the 1/3 power
\begin{equation}
\ds
v_n^{AA}(C=0) = v_n^{PbPb}(C=0) \left(\frac{208}{A} \right)^{1/3}\ .
\label{eq:v-n,A}
\end{equation}
Indeed, the number of ``source-points'' is proportional to the area of 
the overlap region, i.e. $R^2$, and in accordance with ``the 
fluctuation scenario'' given by Eq.(\ref{eq:Tp11}) we get 
\begin{equation}
\ds
v_3 (C) \sim \frac{1}{\sqrt{N_{\rm sources}}} \sim \frac{1}{\sqrt{R^2}} 
\sim \frac{1}{R} \sim A^{-1/3} \ .
\label{v3-A_1/3}
\end{equation}
For lighter colliding system, such as \XeXe, both $v_2$ and $v_3$ will
increase by factor of 1.17 compared to \PbPb central collisions, whereas 
for \CuCu Eq.(\ref{eq:v-n,A}) predicts 1.5 stronger flows. 
ALICE Collaboration in \cite{ALICE_XeXe_18} presented the measurements
of the flow harmonics of charged particles in \XeXe collisions at
$\sqrtsnn = 5.44$~TeV. 
Both $v_2$ and $v_3$, measured in the \XeXe collisions at centralities 
from 3\% to 7\%, are larger than those in \PbPb collisions measured at 
$\sqrtsnn = 5.02$~TeV; see Fig.~2 of \cite{ALICE_XeXe_18}. This is 
qualitatively in line with our predictions. However, only result for
triangular flow is in a perfect agreement with our estimate of the 
increase by 1.17. Elliptic flow in \XeXe collisions at the same 
centralities appears to be 1.35 times stronger than that in \PbPb 
collisions, indicating the influence of the collision geometry. Also,
the elliptic flow in \CuCu collisions at RHIC energies was measured by
the STAR Collaboration in \cite{STAR_v2_CuCu}. Unfortunately, no 
results for triangular flow were presented, and the most central bin
was 0-10\%. Here the role of pure fluctuations in the $v_2$ is 
significantly obscured. It would be interesting to check this 
$A^{-1/3}$ dependence for other colliding systems as well.


\section{Conclusion}
\label{concl}

The role of (i) shape of the overlap area of noncentral nuclear 
collisions and (ii) fluctuations of interacting centers, called {\it 
sources}, in the formation of elliptic and triangular flows are studied 
within the MC hard sphere, MC Glauber, and Magma models. 
Our investigation shows that the third flow harmonic $v_3$ has merely 
fluctuation origin. Its centrality dependence is determined by the
variation of the overlap area with the changing centrality of 
heavy-ion collisions $C$ and is well fitted to the simple ``fluctuation'' 
formula given by Eq.(\ref{eq:Tp11}). In contrast, elliptic flow 
coefficient $v_2$ is closely related to the collision geometry. The 
``fluctuation'' contribution to $v_2$ reveals itself in the most central 
collisions $\sigma/ \sigma_{geo} = 0-2\%$ only. In this centrality 
interval the absolute value of all harmonics is simply determined by the 
number of ``source-points'' and is independent of the shape of the 
overlapping region. Both MC hard sphere and Magma models give quite 
interesting prediction for 
the behavior of magnitude of all harmonics with the variation of atomic 
number $A$. Namely, the magnitude of the signal should scale as 
$A^{-1/3}$ in the most central collisions ($0-2\%$). 
This prediction is in a very good agreement with the experimental 
results obtained by ALICE Collaboration for measurements of $v_3$ in
\XeXe collisions at $\sqrtsnn = 5.44$~TeV and \PbPb collisions at
$\sqrtsnn = 5.02$~TeV \cite{ALICE_XeXe_18}, but underestimates the 
effect for $v_2$ which appears to be 1.35 times stronger in 
semi-central \XeXe collisions. To get rid of additional effects 
caused by collision geometry, here one has to go to very central 
collisions.  
Recall, that
this centrality region is under intensive theoretical and experimental 
study now; see, e.g., recent paper \cite{Zakharov} and references 
therein. The calculated initial eccentricities $\varepsilon_2$ and 
$\varepsilon_3$ can be used as an input to phenomenological models like 
the \hydjet to improve the description of the centrality dependence of 
the flow azimuthal characteristics.


\section*{Acknowledgments}
Fruitful discussions with A.I.~Demyanov, C.~Loizides and I.P.~Lokhtin 
are gratefully acknowledged. 
This work was supported in parts by Russian Foundation for Basic
Research (RFBR) under Grants No. 18-02-40084 and No. 18-02-40085,
and by Norwegian Agency for International Cooperation (SIU) under Grant 
UTF-2016-long-term/10076. 
E.E.Z. acknowledges support of the Norwegian Research Council (NFR) 
under grant No. 255253/F50, ``CERN Heavy Ion Theory."


\end{document}